\documentclass[aps,prb,preprint,groupedaddress]{revtex4-1}
\usepackage{graphicx}

\begin{document}


\title{Saturation and negative temperature coefficient of electrical resistivity in liquid iron-sulfur alloys at high densities from first principles calculations}


\author{Fabian Wagle}
\affiliation{Bayerisches Geoinstitut, Universit\"at Bayreuth, 95440 Bayreuth, Germany.}
\author{Nico de Koker}
\affiliation{School of Geosciences, University of the Witwatersrand, Johannesburg, South Africa.}
\author{Gerd Steinle-Neumann}
\affiliation{Bayerisches Geoinstitut, Universit\"at Bayreuth, 95440 Bayreuth, Germany.}



\begin{abstract}
We report results on electronic transport properties of liquid Fe-S alloys at conditions of planetary cores, computed by first-principle techniques in
the Kubo-Greenwood formalism. We describe a combined effect of resistivity saturation due to temperature, compression, and chemistry by comparing
the electron mean free path from the Drude response of optical conductivity to the mean interatomic distance. At high compression and high sulfur
concentration the Ioffe-Regel condition is satisfied, and the temperature coefficient of resistivity changes sign from positive to negative. We
show that this happens due to a decrease of the $d$-density of states at the Fermi level in response to thermal broadening.
\end{abstract}

\pacs{}

\maketitle

\section{Introduction}

An understanding of the stability of planetary magnetic fields and the thermal evolution of terrestrial planets is closely related to the
characterization of electronic transport properties of liquid Fe and Fe-alloys that make up the dynamo-active portions of their cores. Recent years
have seen significant progress in this direction, and both electrical ($\sigma$) and thermal conductivity ($\lambda_{th}$) have been determined at
high pressure ($P$) and high temperature ($T$) by means of {\it ab-initio} simulations \citep{DeKoker2012,Pozzo2012,Pozzo2013} and
experiments.\citep{Seagle2013,Gomi2013,Gomi2016,Ohta2016,Suehiro2017} While a consensus has emerged that $\sigma$ at conditions of planetary cores is 
significantly higher than previously thought,\citep{Stacey2001,Stacey2007} there is considerable controversy on values of
$\lambda_{th}$\citep{DeKoker2012,Pozzo2012,Pozzo2013,Konopkova2016,Pourovskii2017} 
that includes a discussion on the validity of the Wiedemann-Franz law that relates both electronic transport quantities.

For the Earth's core, Fe is likely alloyed with silicon and/or oxygen\citep{Tsuno2013,Badro2015} that have therefore been the focus of previous
studies.\citep{DeKoker2012,Pozzo2013,Seagle2013,Gomi2013} By contrast, in the cores of Mercury and Mars, sulfur is expected to be the dominant light element alloying with iron:\citep{Hauck2013,Lodders1997} It is cosmically abundant and shows a high solubility in liquid iron due to its compatibility in electronic structure and the similar atomic size of Fe and S.\citep{Alfe1998,Hirose2013} In the Earth's core, sulfur is unlikely to play an important role as the giant Moon-forming impact has probably led to the loss of this moderatly volatile element.\citep{Dreibus1996} 

The observed decrease of conductivity ($\sigma \propto 1/T$) of liquid metals in experiments\citep{VanZytveld1980,Desai1984} and computations, also at
high $P$,\citep{DeKoker2012} is consistent with the Bloch-Gr\"uneisen law for solids above the Debye temperature ($\theta_D$) that describes the shortening of the electron mean free path $x_{\mathrm{eff}} \propto 1/T$. In the quasi-free electron model, scattering events in the liquid occur due to the interaction of electrons with atomic potentials.\citep{Ziman1961} For this scattering mechanism, the interatomic distance sets a lower bound for the mean free path which is known as the Ioffe-Regel condition,\citep{Ioffe1960} leading to saturation. Resistivity saturation has been found to be an important factor in highly resistive transition metals and their alloys,\citep{Gunnarsson2003} in which $x_{\mathrm{eff}}$ is already short, due to the following static and dynamic effects: 

{\it (i)} Experiments at ambient $P$ reveal that a high concentration of impurities can shorten $x_{\mathrm{eff}}$ sufficiently, since the alloying
element introduces compositional disorder.\citep{Mooij1973} Chemically induced saturation continues to take place at high $P$, as has been shown for
the Fe-Si-Ni system.\citep{Gomi2016} \citet{Gomi2016} combined diamond anvil cell experiments with first principles calculations and show that Matthiessen's rule \citep{Ashcroft1976} breaks down close to the saturation limit.

{\it (ii)} Increasing thermal disorder also induces saturation, as has been demonstrated by analyzing the temperature coefficient of resistivity (TCR)
in NiCr thin films.\citep{Mooij1973} Recent computations\citep{Pozzo2016} observe a sub-linear trend of $\rho(T)=1/\sigma$ for hexagonal close packed
(hcp) iron at $P$ of the Earth's inner core.

{\it (iii)} In addition to impurities and $T$, pressure can lead to saturation. This has been shown for the Fe-Si system in the large volume press.\citep{Kiarasi2015}

Since electrical conductivity measurements of liquid iron and its alloys at conditions of the Earth's core are challenging,\citep{Dobson2016} high $P$
studies extrapolate ambient $T$ \citep{Gomi2013,Suehiro2017} or high $T$ experiments \citep{Ohta2016} for the solid to the melting temperature and the
liquid phase, accounting for saturation by a parallel resistor model. The extrapolation of their models supports low values of $\rho$ for the Earth's
core, consistent with computational studies. \citep{DeKoker2012,Pozzo2012,Pozzo2013} Here, we investigate the electronic transport properties for
liquid iron-sulfur alloys based on first principle simulations to complement the existing results for Fe\citep{DeKoker2012,Pozzo2012} and the Fe-O-Si
system,\citep{DeKoker2012,Pozzo2013} and to compare to recent experiments in the Fe-Si-S system.\citep{Suehiro2017} The first principles approach also
provides the opportunity to explore resistivity saturation in terms of the Ioffe-Regel condition and the TCR by means of the electronic structure.


    \section{Methods}
    \label{sec:methods}

We generate representative liquid configurations using density functional theory based molecular dynamics (DFT-MD) simulations, for which we then perform electronic linear repsonse calculations to obtain transport properties. 

    \subsection{Molecular dynamics simulations}

DFT-MD simulation cells contain 128 atoms and the calculations are performed in the $N$-$V$-$T$ ensemble, using the plane-wave code
VASP.\citep{Kresse1993,Kresse1996a,Kresse1996b} Cubic cells in a volume range between 7.09 and 11.82 \AA$^3$/atom (six equally spaced volumes,
covering the $P$-range of the Earth) and sulfur contents of 12.5 (Fe$_7$S) and 25 at.\% (Fe$_3$S) ($\sim$7.6 and $\sim$16 wt.\%) are set up by
randomly replacing Fe atoms in molten configurations from previous simulations.\citep{DeKoker2012} At 8.28 \AA$^3$/atom we also set up Fe$_{15}$S and
Fe$_{27}$S$_5$ compositions to consider the dependence of resistivity on composition in more detail. Atomic coordinates are updated using a time step
of 1 fs, and $T$ is controlled by the Nos\'e thermostat,\citep{Nose1984} with $T$ between 2000 K and 8000 K. At each time step, the electron density
is computed using the projector-augmented-wave (PAW) method \citep{Kresse1999} with the PBE exchange-correlation functional \citep{Perdew1996} and a
plane wave cutoff energy of 400 eV. Electronic states are occupied according to Fermi-Dirac-statistics at $T$ of the thermostat. Brillouin zone
sampling is restricted to the zone center. After equilibration of $P$, $T$ and the total energy ($E$) is achieved (typically after a few hundred fs),
the DFT-MD simulations are continued for at least 15 ps.
    
    \subsection{Resistivity calculations}

The kinetic coefficients in linear response to an electric field $\mathbf E$ and a thermal gradient $\nabla T$ build up the Onsager matrix $\mathcal L_{ij}$ \citep{Onsager1931}
    \begin{eqnarray}
        \mathbf j_{el}&=&\mathcal L_{11}\mathbf E+\mathcal L_{12}\nabla T;\\
        \mathbf j_{th}&=&\mathcal L_{21}\mathbf E+\mathcal L_{22}\nabla T,
    \end{eqnarray}
where $\mathbf j_{el}$ and $\mathbf j_{th}$ are electrical and thermal current densities, respectively. Electrical conductivity and the electronic contribution to thermal conductivity are then
    \begin{equation}
        \sigma=\mathcal L_{11}
    \end{equation}
    and
    \begin{equation}
        \lambda^{el}_{th}=\frac{1}{e^2T}\left(\mathcal L_{22}-\frac{\mathcal L_{12}^2}{\mathcal L_{11}}\right).
    \end{equation}

We extract at least six uncorrelated snapshots from the MD simulations  (i.e., separated by time periods greater than that required for the velocity
autocorrelation function to decay to zero) and compute Kohn-Sham wavefunctions $\psi_k$, their energy eigenvalues $\epsilon_k$ and the cartesian gradients of the Hamiltonian with respect to a shift in wave-vector $\partial\mathcal H/\partial\mathbf k$ using the \textit{Abinit} software package.\citep{Gonze1997,Gonze2009,Torrent2008} From those, the frequency-dependent Onsager matrix elements are calculated with the Kubo-Greenwood equations
    \begin{widetext}
    \begin{equation}
        \mathcal L_{ij}=(-1)^{i+j}\frac{\hbar e^2}{V_{cell}}\sum\limits_{k^\prime,k}[f(\epsilon_{k^\prime})-f(\epsilon_k)]
        \delta(\epsilon_{k^\prime}-\epsilon_k-\hbar\omega)\langle\psi_k|\hat{\mathbf v}|\psi_{k^\prime}\rangle\langle\psi_{k^\prime}|\hat{\mathbf v}|\psi_k\rangle
        (\epsilon_{k^\prime}-\mu_e)^{i-1}(\epsilon_k-\mu_e)^{j-1},\label{eq:KG}
    \end{equation}
    \end{widetext}
as implemented in the \textit{conducti}-module of \textit{Abinit}.\citep{Recoules2005} In equation (\ref{eq:KG}), $\hbar$ denotes the reduced Planck constant, $e$ the elementary charge, $V_{cell}$ the cell volume, $\omega$ the frequency of the external field, $\hat{\mathbf v}=1/\hbar \cdot \partial\mathcal H/\partial\mathbf k$ the velocity operator and $\mu_e$ the electronic chemical potential.
    
By fitting the Drude formula for optical conductivity
    \begin{equation}
        \Re[\sigma(\omega)]=\frac{\sigma_0}{1+(\omega\tau)^2}\label{eq:drude} 
    \end{equation}
to the Kubo-Greenwood results for each snapshot, we extract the DC limit of conductivity $\sigma_0$ (used without subscript elsewhere) and 
effective relaxation time $\tau$. Thermal conductivity is extrapolated linearly to the limit $\omega\rightarrow0$ over a $\hbar\omega$-range of 2 eV. We
average $\sigma$, $\tau$ and $\lambda_{th}$ over the snapshots and take one standard deviation as uncertainty. Calculations with denser
grids of $2\times2\times2$ and $3\times3\times3$ $k$-points show that $\sigma(\omega)$ is sufficiently converged (to within 3\%) in calculations using
a single $k$-point ({\it cf.} Figure S1 in the Supplemental Material). 

Resulting $\rho(V,T)$ and $\lambda_{th}(V,T)$ are fit with a physically-motivated closed expression (Appendix A) to interpolate between results and extrapolate to conditions not investigated.

    \subsection{Electron density of states}

We compute the site-projected and angular momentum-decomposed electron densities of states (DOS) by the tetrahedron method,
\citep{Jepsen1971,Lehmann1972} using a non-shifted $2\times2\times2$ $k$-point grid with small energy increments of $1.4\cdot10^{-3}$ eV. Radii of the
atomic spheres, in which the angular-momentum projections are evaluated, have been chosen to be space filling and proportional to the radii of the
respective PAW-spheres.\citep{Kresse1999} The DOS is computed for the same snapshots as those used for the evaluation of the Kubo-Greenwood equations, and re-binned with an energy window of $\sim 1/2 \cdot k_B T$ to resolve $T$-dependent features in the vicinity of the Fermi energy ($E_F$). This results in a strongly varying DOS which is independent of the smearing parameter.


\section{Results and discussion}
\label{sec:res}
\subsection{Electrical resistivity}
    \label{sec:rho}
For the low impurity composition Fe$_7$S, we find a dependence of $\rho$ on $V$ and $T$ similar to that predicted in previous studies on pure Fe,
Fe-Si and Fe-O systems \citep{DeKoker2012} (Figure \ref{fig:rhoT}, Tables S1 and S2 in the Supplemental Material). Resistivity increases with $V$ and $T$ and can be reasonably well described by a linear $T$-dependence above $\Theta_D$ ($\sim 1000$ K at low compression based on the equation of state parameters, {\it cf.} Appendix B and Table S3 in the Supplemental Material), consistent with Bloch-Gr\"uneisen theory. With decreasing $V$, $\Theta_D$ increases based on the thermodynamic parameters from our DFT-MD simulation, and values for $\rho$ decrease. This behavior is well captured with the resistivity model of Appendix A.

%
%
\begin{figure}[htbp]
\centering
    \includegraphics[width=\linewidth]{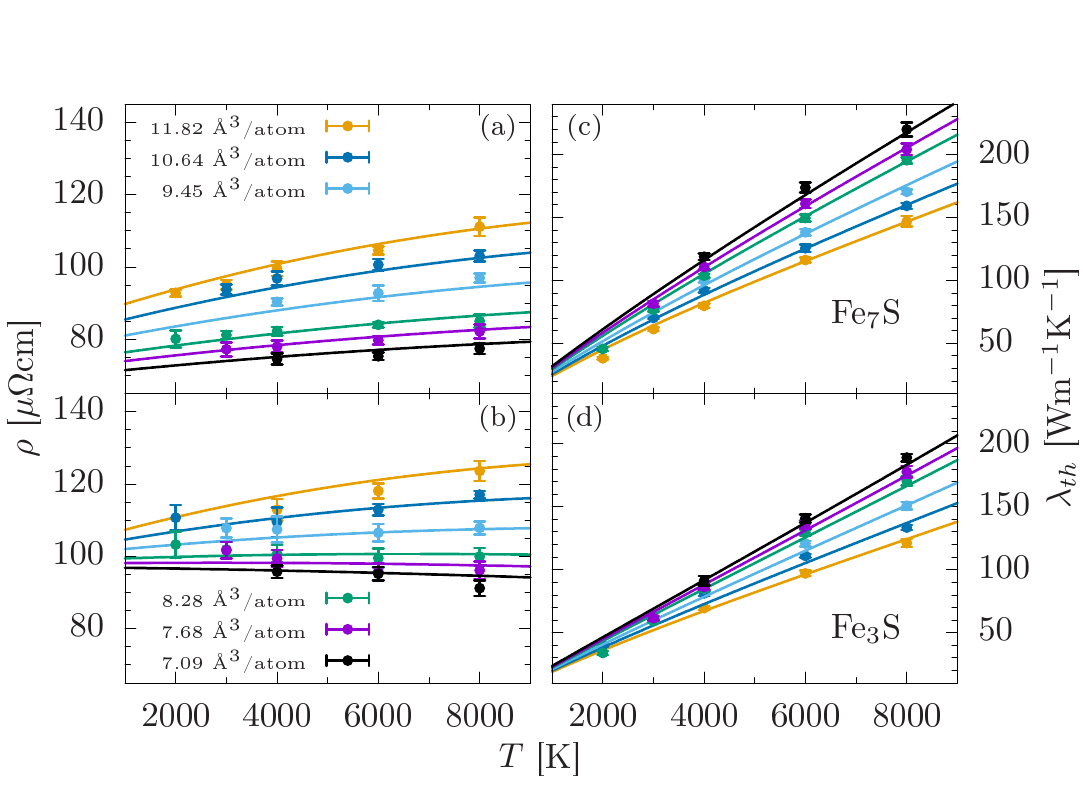}
\caption{Electronic transport properties of liquid Fe--S alloys as a function of temperature. The temperature coefficient of electrical resistivity of liquid Fe$_7$S (a) and
    Fe$_3$S (b) decreases with compression due to saturation. For Fe$_3$S, the temperature coefficient of resistivity becomes negative along the smallest $V$ isochores. Solid
    lines represent the best fit of equation (\ref{eq:fit}) to resistivity. Panels (c) and (d) show the electronic contribution to thermal conductivity of liquid Fe$_7$S and 
    Fe$_3$S respectively. Solid lines have been calculated from the best fits to $\rho(V,T)$ (equation \ref{eq:fit}) and the effective Lorenz number $L(V,T)$ (equation
    \ref{eq:lorenz}). Tabulated values for $\rho$, $\lambda_{th}$ and $L$ are given in Tables S1 and S2 in the Supplemental Material.}
\label{fig:rhoT}
\end{figure}
%

Absolute resistivities for both compositions in the Fe-S system are similar to those for Fe-Si with the same light element
concentration,\citep{DeKoker2012} and higher than those for pure Fe and in the Fe-O system.\citep{DeKoker2012,Pozzo2012} This is in contrast to
experimental work\citep{Suehiro2017} that estimated $\rho$ for the solid phase in a ternary Fe-Si-S system and calculated the S impurity resisitvity
by using Matthiesen's rule based on previous experimental results for Fe \citep{Ohta2016} and Fe-Si.\citep{Gomi2016} \citet{Suehiro2017} find that the
influence of S on resistivity is significantly smaller than that of Si.\citep{Gomi2016} The experiments had to rely on this indirect determination of resistivity
reduction due to sulfur, as S is hardly soluble in solid Fe at ambient $P$ and it is therefore difficult to synthesize a homogeneous phase as a starting
material in experiments.\citep{Li2001,Stewart2007,Kamada2012,Mori2017} Further, Matthiesen's rule, applied in the analysis of the data, does not hold for systems with saturated resistivity.\citep{Gomi2016} 

For higher sulfur concentration, we find that $\rho$ increases (Figure \ref{fig:rhoT}, {\it cf.} Figure S2 in the Supplemental Material) and that the Bloch-Gr\"uneisen behavior breaks down. The temperature coefficient of resistivity decreases with compression, up to the extreme case where it changes sign and becomes negative for Fe$_3$S at the smallest two volumes we consider.

Negative TCR have been observed for liquid and amorphous solid metals, for which the maximum momentum change of a scattered electron $2k_F$ falls in
the region close to the principle peak of the structure factor $S(q)$, as in case of metals with two valence electrons, e.g., Eu, Yb and Ba with a
6$s^2$ valence configuration,\citep{Guntherodt1976} and Cu-Zr metallic glasses.\citep{Waseda1978} It is one of the great successes of Ziman theory for
the resistivity of liquid metals \citep{Ziman1961,Faber1965} to explain the negative TCR in these systems. Ziman theory can not account for the
negative TCR that we predict for Fe$_3$S at high compression. As for iron and the other Fe-alloys considered by \citet{DeKoker2012}, $2k_F$ is near the first minimum in $S(q)$ (Figure S3 in the Supplemental Material), thermal broadening of the structure factor will lead to positive TCR over the entire compression range. This suggests that the negative TCR is a secondary effect, driven by changes in electronic structure (Section \ref{sec:edos})  that is only noticable once resistivity saturation is reached by compression and impurities simultaneously.

\subsection{Mean free path}
\label{sec:mfp}
\begin{figure}[htbp]
\centering
    \includegraphics[width=.8\linewidth]{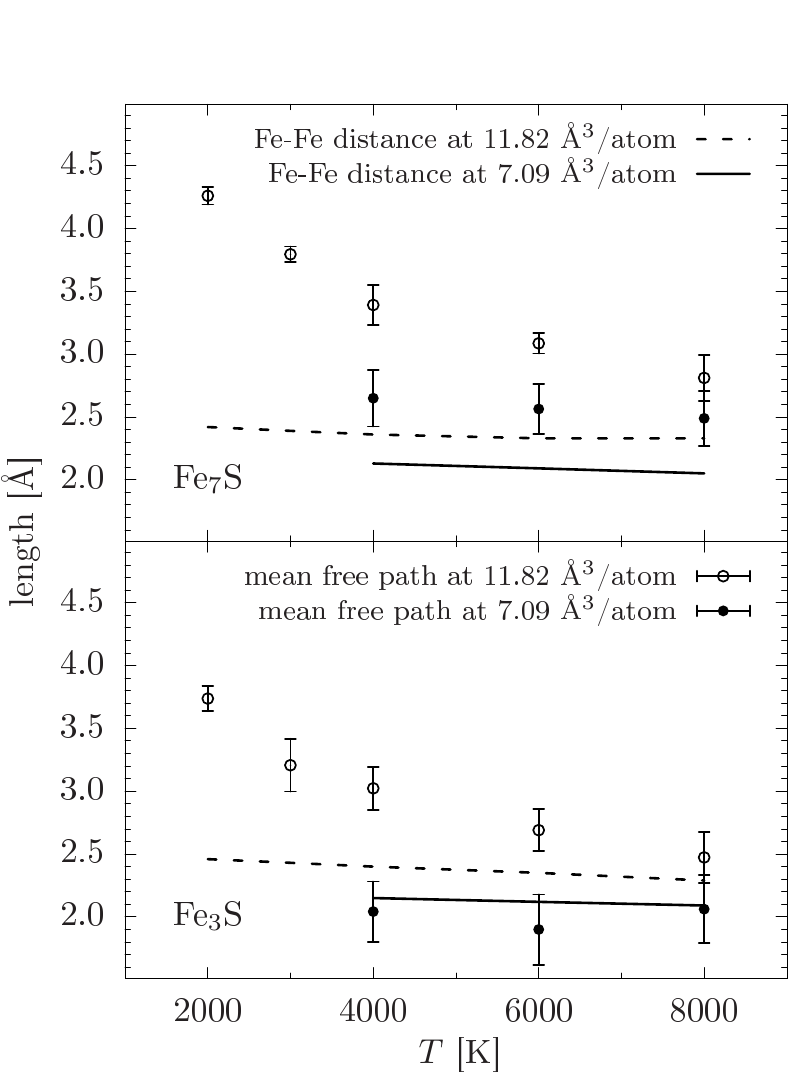}
\caption{Electron mean free path for liquid Fe$_7$S (top) and Fe$_3$S (bottom) for two cell volumes (near ambient $P$ and largest compression) as a function of temperature, 
    obtained by a Drude-fit to optical conductivity (equation \ref{eq:drude}). The mean free path approaches the interatomic distance (solid lines, first peak position of the 
    partial radial distribution function) with increasing compression and impurity concentration. For Fe$_3$S at the smallest cell volume, the Ioffe-Regel condition is reached.}
\label{fig:mfp}
\end{figure}

In order to understand the effect of resistivity saturation from a semi-classical picture of electron transport, we calculate the effective electron mean free path as $x_{\mathrm{eff}}=v_F\tau$, where $v_F=(\hbar/m)\cdot(3\pi^2n_{\mathrm{eff}})^{1/3}$ is the Fermi velocity, $n_{\mathrm{eff}}=(m\sigma_0)/(e^2\tau)$ the effective number density of conduction electrons and $m$ the electron mass. Figure \ref{fig:mfp} reveals three distinctive features:

{\it (i)} For ambient $P$ volumes ($V=11.82$ \AA$^3$/atom), $x_{\mathrm{eff}}$ approaches the mean interatomic distance asymptotically with increasing $T$, consistent with dynamic resistivity saturation.\citep{Mooij1973,Pozzo2016} 

{\it (ii)} At the lowest cell $V$ considered ($V=7.09$ \AA$^3$/atom), the $T$-dependence of $x_{\mathrm{eff}}$ vanishes within uncertainty. In addition,  $x_{\mathrm{eff}}$  becomes shorter than at lower compression due to the increased density of scattering centers. At first glance, this observation appears to be inconsistent with the fact that $\rho$ decreases with compression, but can be understood in terms of electronic structure (Section \ref{sec:edos}). 
    
{\it (iii)} With increasing sulfur concentration, $x_{\mathrm{eff}}$ decreases significantly. This reflects the expected behavior of an increased probability of impurity-caused scattering. 

For the highest compression the Ioffe-Regel condition is reached for Fe$_3$S as $x_{\mathrm{eff}}$ becomes equal to the mean interatomic distance within uncertainty.

    \subsection{Electronic structure}
    \label{sec:edos}

Most of the electric current in transition metals is transported by $s$-electrons, which can scatter into $d$-states with a far lower Fermi velocity.\citep{Mott1936} Partially filled $d$-bands with a high DOS at the Fermi level lead to a high probability of $s$-$d$ scattering events, which dominate resistivity over $s$-$s$ processes.\citep{Mott1972} 

Site-projected and angular momentum-decomposed densities of states (LDOS) show similar changes in response to compression and $T$ (Figures S4 and S5
of the Supplemental Material). Generally, peaks broaden and the Fe $d$-LDOS at $E_F$ decreases, resulting in fewer states available for $s$-electrons to scatter into.
The response of the electronic structure to compression is a dominant feature as dispersion of electronic bands increases significantly due to
stronger interactions (Figure S4 in the Supplemental Material).\citep{Cohen1997} 

For increasing $T$, changes in the DOS are less pronouced (Figure S5 in the Supplemental Material) and reflect dynamic short range changes in the liquid structure that can lead to smaller interatomic distances\citep{Hunt2003} that is also expressed by thermal pressure.\citep{Pozzo2016} This is a small effect, and the negative TCR can only be observed when compression and chemical saturation in the system has been reached.

Electronic states of iron dominate the DOS of the liquid Fe-S alloys near $E_F$. The densities of states for Fe and Fe$_3$S are quite similar
at the same $V$ and $T$ (Figures S4 and S5 in the Supplemental Material) and the broadening in the vicinity of $E_F$ due to compression and $T$, respectively, is almost 
identical. Therefore, sulfur contributes to the overall resistivity behavior in the Fe-S systems only by shortening $x_{\mathrm{eff}}$ through impurity scattering 
as discussed in Section \ref{sec:mfp} (Figure \ref{fig:mfp}). In comparison to silicon and oxygen, sulfur appears to be more efficient in doing so due to its similar atomic size and the 
efficient bonding with iron, resulting in high Fe-S coordination numbers. \citep{Alfe1998}
    
 \subsection{Thermal conductivity}
 \label{sec:lambda}

Since lattice vibrations play only a minor role in heat transport through metals, the electronic
contribution to thermal conductivity $\lambda^{el}_{th}$ represents total conductivity $\lambda_{th}$ to a good approximation. \citep{Ashcroft1976} Similar to the results
for $\rho$, we find the Kubo-Greenwood values for $\lambda_{th}$ (Figure \ref{fig:rhoT}) to be consistent with the ones of
liquid Fe-Si alloys, and somewhat larger than those of Fe-O liquids from previous computations with the same light element
concentrations.\citep{DeKoker2012} Contrary to electrical resistivity, we do not see any sign of saturation in $\lambda_{th}$, putting the validity of
the Wiedemann-Franz law with a constant value of the Lorenz number $L_0\approx2.44$ W$\Omega$/K$^2$ from Drude-Sommerfeld theory in question. 
Indeed, thermal conductivity is significantly overestimated by using $L_0$ and the the resistivity model (Appendix A) compared to 
the values computed directly with the Kubo-Greenwood equations (equation \ref{eq:KG}).

Recently, electron-electron scattering has been suggested to contribute significantly to $\lambda_{th}$ of hcp iron at high $P$, but not to 
$\rho_{el}$,\citep{Pourovskii2017} an effect that is ignored in the independent electron approximation of the Kubo-Greenwood approach. 
However, it remains an open question to what degree this contribution affects thermally disordered systems. Electronic transport critically depends on the
electronic structure at the Fermi level, which is quite different for a high density liquid at high $T$, compared to a perfect crystal.
Until the influence of electron-electron scattering on transport properties of disordered $3d$ transition metals and their alloys is better understood, 
values for $\lambda_{th}$ from the Kubo-Greenwood approach should be used with caution.

    \subsection{Application to planetary interiors}
    \label{sec:planets}
\begin{figure}[htbp]
\centering
    \includegraphics[width=.8\linewidth]{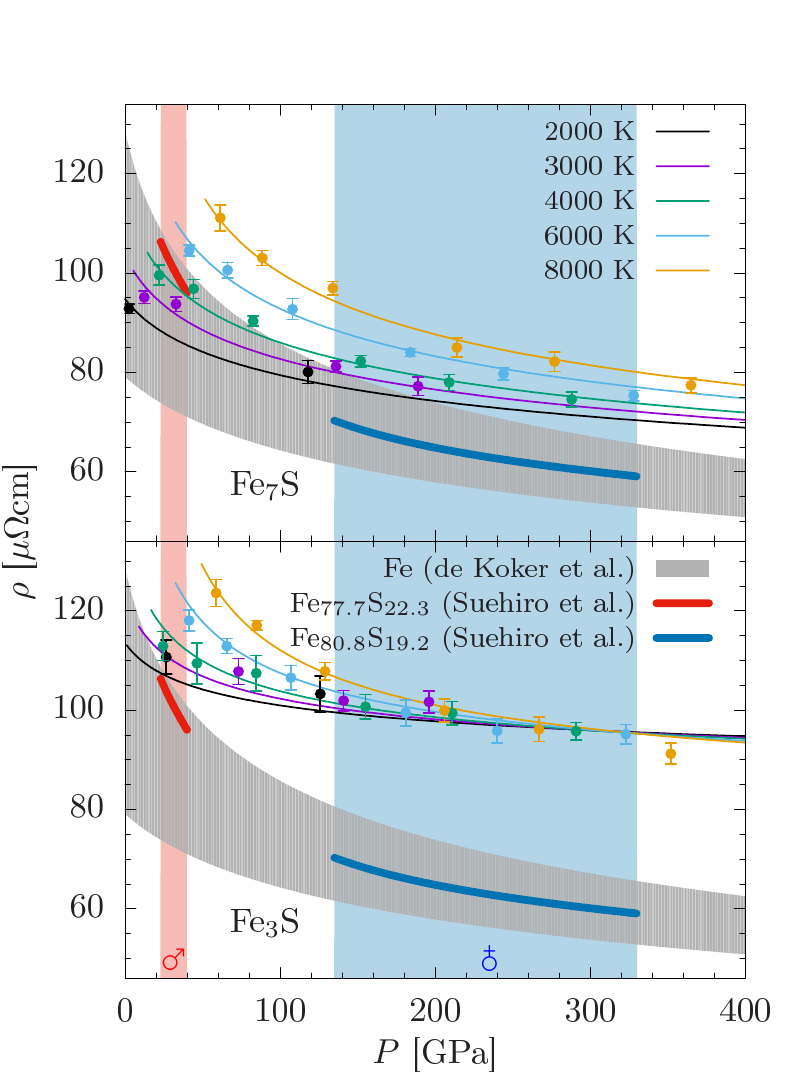}
\caption{Electrical resistivity of liquid Fe--S alloys as a function of pressure for Fe$_7$S (top) and Fe$_3$S (bottom). Solid lines are best fits of
    a parallel resistor model to $\rho(V,T)$ (equation \ref{eq:fit}) converted from $V$-$T$ to $P$-$T$ conditions using the equation of state fits
    (Appendix \ref{sec:EOS}). Results from an experimental study \citep{Suehiro2017} along a model areotherm (red line) and geotherm (blue line) as well 
    as computational results for pure Fe \citep{DeKoker2012} between 2000 and 8000 K (gray area) are included for comparison.}
\label{fig:rhoP}
\end{figure}

We convert resistivity values and fits in $V$-$T$ space (Appendix \ref{sec:model} and Table \ref{tab:fit}) to $\rho(P,T)$ by using the
self-consistently obtained equations of state for Fe$_7$S and Fe$_3$S (Appendix \ref{sec:EOS}, Figure S6 and Table S3 in the Supplemental Material). Resistivity values for Fe$_7$S and Fe$_3$S (Figure \ref{fig:rhoP}) are substantially larger than the corresponding ones for pure iron. While resistivities for Fe$_7$S along different isotherms continue to show distinctive $P$-trends, they become indistinguishable for Fe$_3$S at high $P$ due to the combined saturation effects discussed in Section \ref{sec:mfp}. For Fe$_3$S, resistivity saturates at $\sim$100 $\mu\Omega$cm, a value which remains approximately constant and $T$-independent over the $P$-range of the Earth's outer core, similar to the behavior of Fe$_3$Si.\citep{DeKoker2012}

There is a large discrepancy between our results and the high $T$ extrapolation of experimental resistivity,\citep{Suehiro2017} reported along model adiabats
in the cores of Mars and the Earth.\citep{Fei2005,Kamada2012} Despite the similar composition between the work presented here and the experiments (that fall
between Fe$_3$S and Fe$_7$S, towards the higher sulfur concentration), the experimental profile for Earth's core shows significantly lower values,
more consistent with the Kubo-Greenwood results for pure Fe.\citep{DeKoker2012,Pozzo2012} 
Model values of \citet{Suehiro2017} in the $P$-range of the Martian core are closer to our results (Figure \ref{fig:rhoP}), but the slope $(\partial \rho/\partial P)_S$ in the model based on experiments is significantly larger than in our work.

A small contribution to the difference between the experimental data and our results may come from the fact that the experiments have been performed
for the solid and the simulations on the liquid, and resistivity increases discontinuously across the melting point for metals and their alloys both at
ambient\citep{Rosenfeld1990} and high $P$.\citep{Secco1989,Silber2017,Ezenwa2017a,Ezenwa2017b} However, based on the Ziman
approximation,\citep{Ziman1961} this difference is expected to decrease with $P$ if density and compressibility of the coexisting solid and liquid
phases become more similar. For pure iron, for example, this discontinuity is likely to become negligible at conditions of the Earth's
core.\citep{Wagle2018a} 
Rather than the difference decreasing with $P$ as expected, it increases between the experimental data\citep{Suehiro2017} and our computational results (Figure \ref{fig:rhoP}).


\section{Conclusions}

We present electronic transport properties of liquid Fe-S alloys from DFT-MD simulations at conditions relevant for the cores of terrestrial planets. We find absolute values of electrical resistivity and thermal conductivity to be consistent with those of other Fe-light element alloys reported in previous work, \citep{DeKoker2012,Pozzo2014} ranging from 75 to 125 $\mu\Omega$cm and 30 to 220 Wm$^{-1}$K$^{-1}$. Fe alloys with low S content exhibit a positive TCR along isochores, which gradually decreases upon compression. We show that this is due to a compression-induced resistivity saturation by comparing the electron mean free path to interatomic distances. For high S concentrations (Fe$_3$S), the mean free path is further shortened by increased impurity scattering, sufficient to reach the Ioffe-Regel condition at the lowest volumes, resulting in a saturation of resistivity. At these conditions the TCR becomes negative which is caused by a  decrease of the Fe $d$-density of states at the Fermi level.

For applications in planetary physics, we provide models for $\rho(V,T)$ and $\lambda_{th}(V,T)$ (Appendix A), which, in combination with a self-consistent thermodynamic equation of state (Appendix B), can be translated to $P$-$T$ conditions of planetary cores. 
\appendix
    \section{Model for electrical and thermal conductivity}
    \label{sec:model}
    We describe the resistivity behavior $\rho(V,T)$ by a parallel resistor model:
    \begin{equation}
        \frac{1}{\rho(V,T)}=\frac{1}{\rho_{BG}(V,T)}+\frac{1}{\rho_{sat}(V)}+\frac{1}{\rho_{el}(T)},\label{eq:fit}
    \end{equation}
    where 
    \begin{equation}
        \rho_{BG}=\rho_0\left(\frac{V}{V_0}\right)^a+\rho_1\left(\frac{V}{V_0}\right)^b\frac{T}{T_0}\label{eq:rhoBG}
    \end{equation}
    is the empirical expression used by \citet{DeKoker2012} based on the Bloch-Gr\"uneisen formula.
    \begin{equation}
        \rho_{sat}=c\left(\frac{V}{V_0}\right)^{\frac{1}{3}}\label{eq:rhosat}
    \end{equation}
    is a term accounting for resistivity saturation and  
    \begin{equation}
        \rho_{el}=d\frac{T_0}{T}\label{eq:rhoel}
    \end{equation}
    describes the effect of thermal broadening of the DOS. The assumptions entering equations (\ref{eq:fit})--(\ref{eq:rhoel}) are: 

    \textit{(i)} Sources of resistivity contributions in equation (\ref{eq:fit}) are independent and therefore conductivities are additive.
    
    \textit{(ii)} In the limit of high $T$, the Bloch-Gr\"uneisen formula is linear in $T$. Both residual resistivity (first term in equation \ref{eq:rhoBG}) and the material
    dependent prefactor of the second term are well described by a power law in $V/V_0$. 
    
    \textit{(iii)} Saturation resistivity (equation \ref{eq:rhosat}) is proportional to interatomic distance and therefore increases 
    $\propto\left(V/V_0\right)^{1/3}$. This is consistent with saturation resistivities for pure Fe reported by \citet{Ohta2016} 
    
    \textit{(iv)} Since the effect of thermal broadening on the DOS at $E_F$ can be attributed to a resistivity contribution due to thermal pressure 
    (Figure S5 in the Supplemental Material), we describe $\rho_{el}$ in equation (\ref{eq:rhoel}) as inversely proportional to $T$.
    
Rather than fitting a model for $\lambda_{th}$ directly, we compute an effective Lorenz number $L$ at each simulation and fit the $L(V,T)$ as\citep{DeKoker2012} 
    \begin{equation}
        L(V,T)=L_R\left(\frac{V}{V_0}\right)^{e}\left(\frac{T}{T_0}\right)^{f}.\label{eq:lorenz}
    \end{equation}
Fit parameters are listed in Table \ref{tab:fit}.
 
\begin{table}
    \caption{Fit parameters of the models for $\rho(V,T)$ (equations \ref{eq:rhoBG}--\ref{eq:rhoel}) and $L(V,T)$ (equation \ref{eq:lorenz}) for liquid Fe, Fe$_7$S and Fe$_3$S. Uncertainties of the fit parameters are large and exceed their values in most cases.}
\label{tab:fit}
\centering
\begin{ruledtabular}
\begin{tabular}{llccc}
                &                   &   Fe    & Fe$_7$S & Fe$_3$S \\
    \hline
    $\rho_{0R}$ & [$\mu\Omega$cm]   &  75.10  &  89.03  &  105.2  \\
    $\rho_{1R}$ & [$\mu\Omega$cm]   &  21.48  &  12.73  &  12.06  \\
    $a$         &                   &  0.792  &  0.389  &  0.124  \\
    $b$         &                   &  1.479  &  1.804  &  2.686  \\
    $c$         & [$\mu\Omega$cm]   &  747.2  &   2077  &   6609  \\
    $d$         & [$\mu\Omega$cm]   &  1405   &   2829  &   2910  \\
    \hline
    $L_R$       & [W$\Omega$/K$^2$] &  2.005  &  2.105  &  1.991  \\
    $e$         &                   & -0.097  & -0.106  & -0.228  \\
    $f$         &                   &  0.041  & -0.027  & -0.022  \\
\end{tabular}
\end{ruledtabular}
\end{table}
    
    \section{Equation of state model}
    \label{sec:EOS}
In order to describe electronic transport properties as a function of $P$, suitable for comparison to experiments and for applications in planetary
models, we fit a thermodynamic model to the Fe$_7$S and Fe$_3$S results that is based on an separation of the Helmholtz energy in an
ideal gas, electronic and excess term.\citep{DeKoker2009,Vlcek2012} The volume dependence of the excess term is represented by Eulerian finite strain
($f$) with exponent $n=2$ and a similarly reduced $T$-term ($\Theta$) with exponent $m=0.79$ and expansion orders $\mathcal{O}_f=3$ and
$\mathcal{O}_{\Theta}=2$, parameters that describe the results for liquid iron well.\citep{DeKoker2012} Figure S6 in the Supplemental Material shows the
quality of the fit for $E$, $P$ and electronic entropy $S_{el}$ of the DFT-MD results. Thermodynamic parameters at reference conditions are summarized in Table
S3 of the Supplemental Material.
\nocite{Nordheim1928}

\begin{acknowledgments}
This work was supported by Deutsche Forschungsgemeinschaft (German Science Foundation, DFG) in the Focus Program ``Planetary Magnetism'' (SPP 1488) 
with grant STE1105/10-1 and Research Unit ``Matter under Planetary Interior Conditions'' (FOR 2440) with grant STE1105/13-1. 
Computing and data resources for the current project were provided by the Leibniz Supercomputing Centre of the Bavarian Academy of Sciences and
the Humanities (www.lrz.de).
We greatly acknowledge informative discussions with Vanina Recoules and Martin Preising on the electron density of states evaluation and helpful comments
by an anonymous reviewer.
\end{acknowledgments}

%

\end{document}



\title{\textit{Supplemental Material:} Saturation and negative temperature coefficient of electrical resistivity in liquid iron-sulfur alloys at high densities from first principles calculations}


\author{Fabian Wagle}
\affiliation{Bayerisches Geoinstitut, Universit\"at Bayreuth, 95440 Bayreuth, Germany.}
\author{Nico de Koker}
\affiliation{School of Geosciences, University of the Witwatersrand, Johannesburg, South Africa.}
\author{Gerd Steinle-Neumann}
\affiliation{Bayerisches Geoinstitut, Universit\"at Bayreuth, 95440 Bayreuth, Germany.}


\date{\today}

\pacs{}

\maketitle

\begin{figure}[htbp]
\centering
    \includegraphics[width=\linewidth]{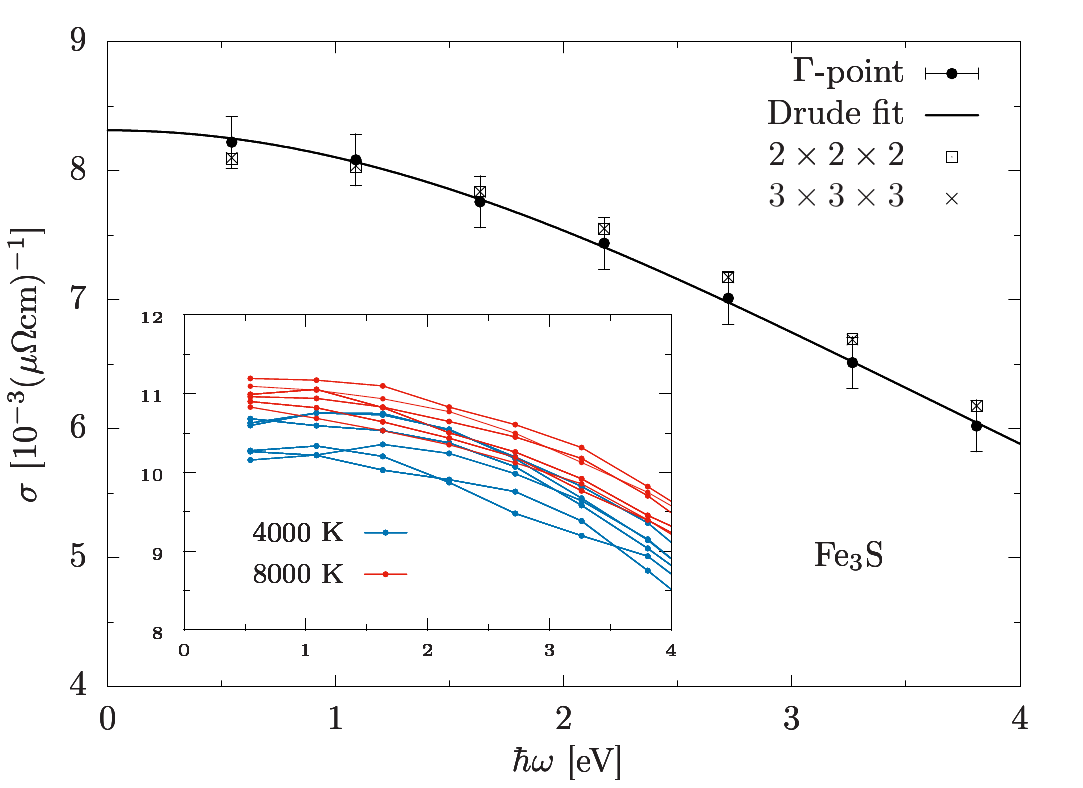}
    \caption{Optical conductivity $\sigma(\omega)$ for Fe$_3$S ($V=11.82$ \AA$^3$/atom, $T=8000$ K) fitted with a Drude model (equation 6). Results
    are converged to within 3\% with respect to Brillouin-zone sampling by using the zone center only. The inset shows various $\sigma(\omega)$ curves
    from different MD snapshots for Fe$_3$S at $V=7.09$ \AA$^3$/atom for 4000 K and 8000 K, demostrating that the negative temperature coefficient 
    of resistivity is statistically significant.}
\label{fig:sigma}
\end{figure}

\begin{figure}[htbp]
\centering
    \includegraphics[width=\linewidth]{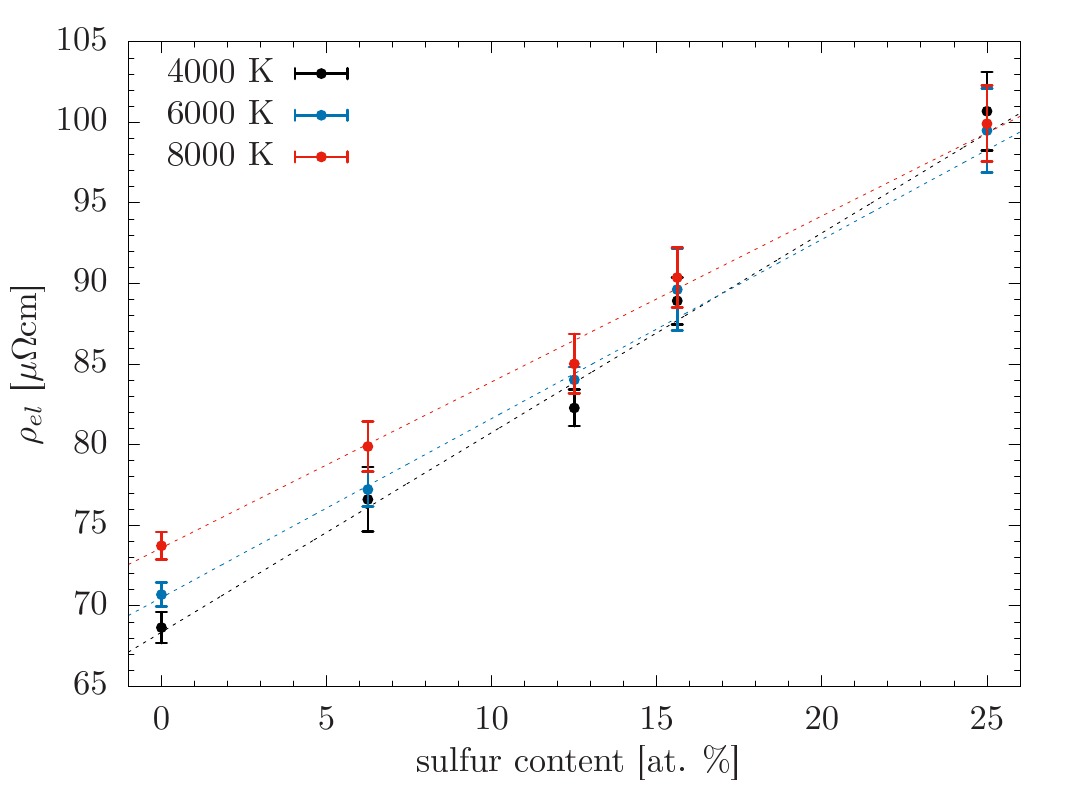}
    \caption{Electrical resistivity as a function of sulfur concentration at a volume of 8.28 \AA$^3$/atom for Fe, Fe$_{15}$S (3.7 wt.\%), Fe$_7$S
    (7.6 wt.\%), Fe$_{27}$S$_5$ (9.6 wt.\%) and Fe$_3$S (16 wt.\% sulfur) with linear regressions to guide the eye. Although the simulations have been
    performed at the same atomic volume, the molar volumes vary along the $x$-axis and are therefore not comparable \textit{sensu stricto} and cannot
    be fit with a Nordheim rule-like expression\footnote[0]{$^{64}$L. Nordheim, Naturwiss. \textbf{16}, 1042 (1928)}.$^{64}$}
\label{fig:rhoC}
\end{figure}

\begin{figure}[htbp]
\centering
    \includegraphics[width=.8\linewidth]{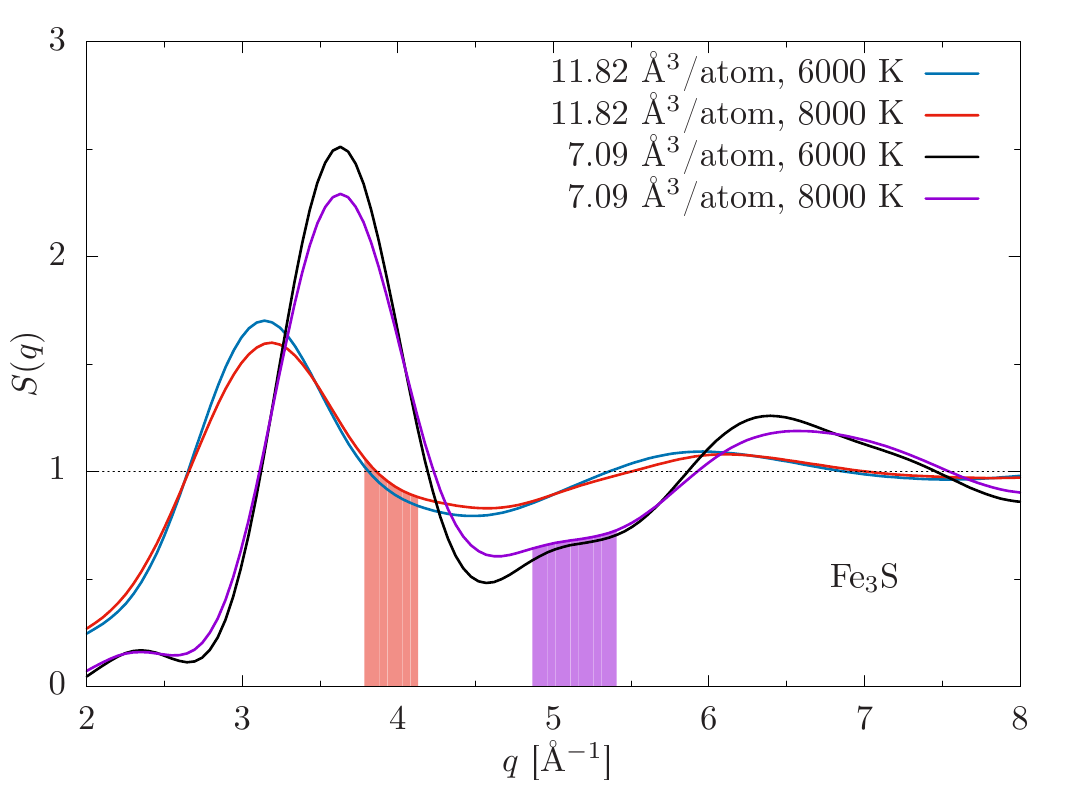}
    \caption{Temperature dependence of static structure factors $S(q)$ of Fe$_3$S for lowest ($V=11.82$ \AA$^3$/atom) and highest compression ($V=7.09$
    \AA$^3$/atom). The shaded areas correspond to the wavenumber of a backscattering event ($2k_F$) at the respective volume, computed by $k_F=(3\pi^2n_\mathrm{eff})^{1/3}$ with
    the corresponding uncertainty. According to Ziman's formula,$^{22}$\footnote[0]{$^{22}$J. M. Ziman, Philos. Mag. \textbf{6}, 1013 (1961)}
    the change of $S(q)$ at this value determines the temperature coefficient of resistivity. Based on the structure factor and the location of $2k_F$, 
    one expects a positive temperature coefficient of resistivity for both compressions.}
\label{fig:sq}
\end{figure}

\begin{figure}[htbp]
\centering
    \includegraphics[width=.8\linewidth]{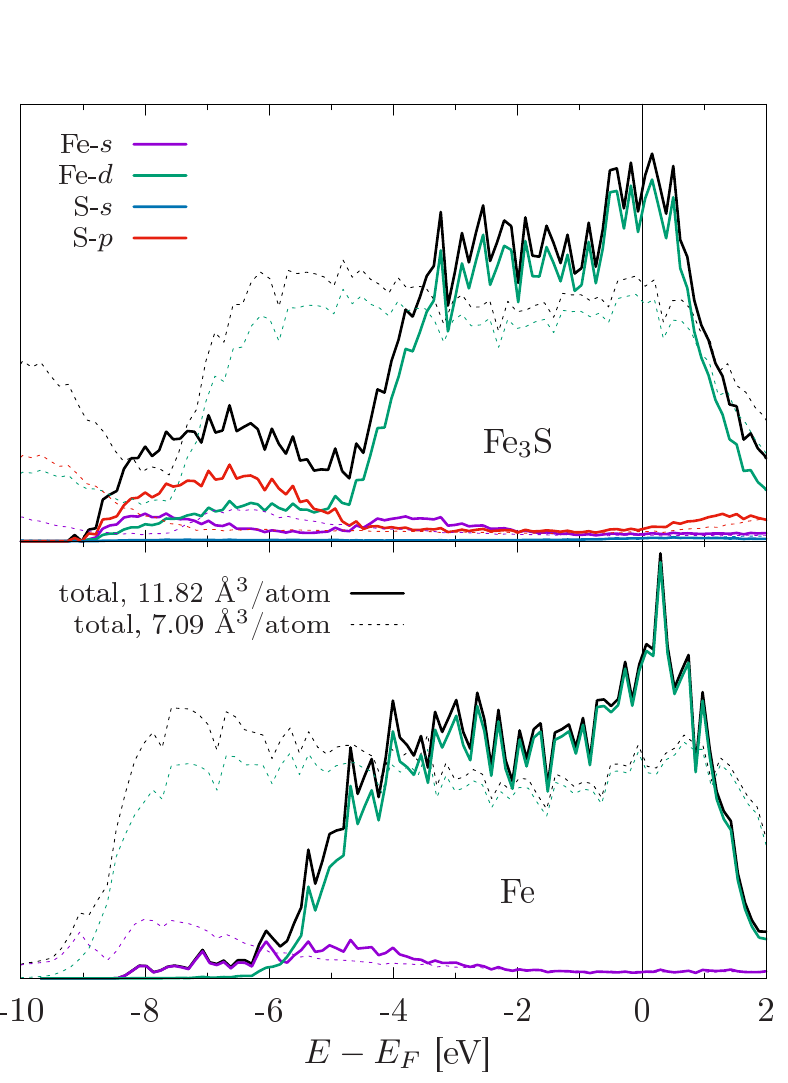}
    \caption{Broadening of site- and angular momentum-projected electron densities of states for Fe$_3$S (top) and pure Fe (bottom) at 4000 K with decreasing cell volume.}
\label{fig:PDOS}
\end{figure}

\begin{figure}[htbp]
\centering
    \includegraphics[width=.8\linewidth]{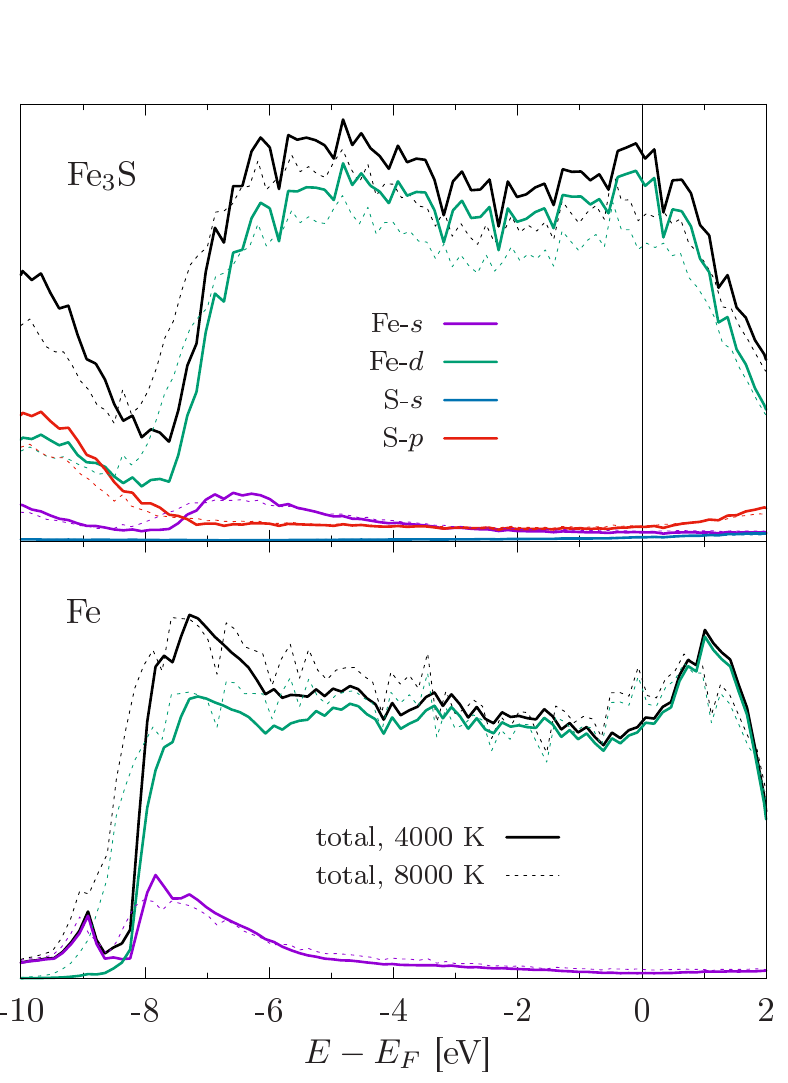}
    \caption{Broadening of site- and angular momentum-projected densities of states with increasing $T$ (4000 K and 8000 K) for Fe$_3$S (top) and pure Fe (bottom) at a cell volume of 7.09 \AA$^3$/atom.}
\label{fig:TDOS}
\end{figure}

\begin{figure}[htbp]
\centering
\includegraphics[width=\linewidth]{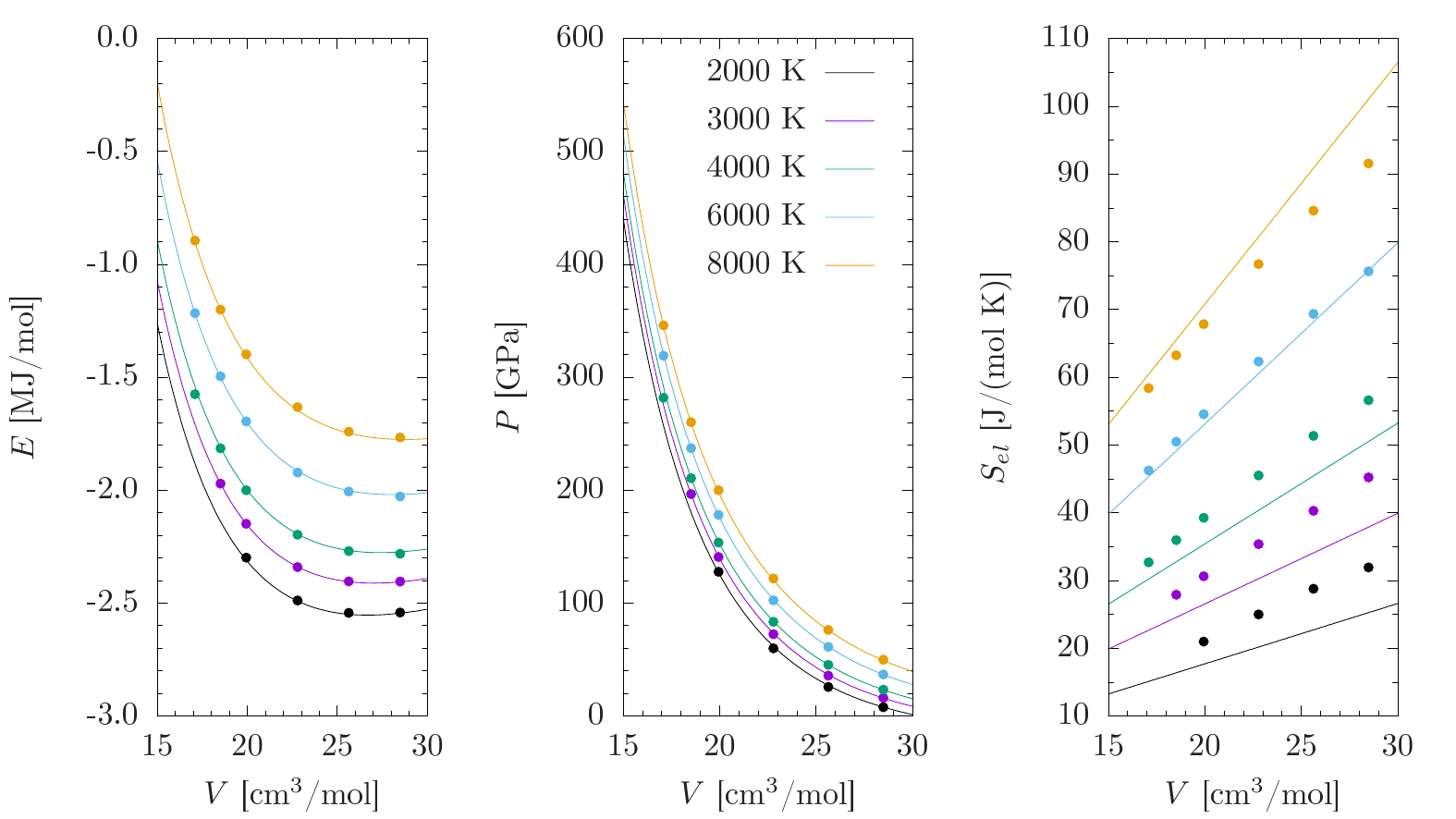}
    \caption{Equation of state fit (lines) to DFT-MD results (dots) for liquid Fe$_3$S by a self consistent thermodynamic
    description:$^{62}$\footnote[0]{$^{62}$N. de Koker and L. Stixrude, Geophys. J. Int. \textbf{178}, 162 (2009)} Internal energy $E$, pressure
    $P$ (without electronic entropy contribution) and electronic entropy $S_{el}$ as a function of volume along isotherms. 
    $S_{el}$ is not particularly well represented by the global fit, but contributes only little to total $P$. The maximal error for $T=8000$ K is well below 2 GPa.}
\label{fig:EOS}
\end{figure}

\begin{table}[htbp]
    \caption{Calculated values for pressure, electrical resistivity, thermal conductivity and Lorenz number of Fe$_{7}$S from first principles
    computations, with reference volume $V_0=11.82$ \AA$^3$/atom. Uncertainties of $P$ due to the equation of state fit are below 1 GPa.}
    \label{tab:cond7}
\centering
\begin{ruledtabular}
\begin{tabular}{cccccc}
    $V/V_0$              & $T$ [K] & $P$ [GPa] & $\rho$ [$\mu\Omega$cm] & $\lambda_{th}$ [Wm$^{-1}$K$^{-1}$] & $L$ [10$^{-8}$W$\Omega$K$^{-2}$] \\
\hline
\multirow{5}{*}{1.0} & 2000    & 3         &  $93\pm1 $             &  $38\pm1 $                    & $1.77\pm0.05$                    \\ 
                     & 3000    & 12        &  $95\pm1 $             &  $62\pm1 $                    & $1.94\pm0.05$                    \\ 
                     & 4000    & 22        &  $100\pm2$             &  $80\pm2 $                    & $1.98\pm0.06$                    \\ 
                     & 6000    & 41        &  $105\pm1$             &  $116\pm2$                    & $2.02\pm0.04$                    \\ 
                     & 8000    & 62        &  $111\pm3$             &  $147\pm4$                    & $2.04\pm0.07$                    \\ 
\hline
\multirow{4}{*}{0.9} & 3000    & 33        &  $94\pm1 $             &  $70\pm1 $                    & $2.19\pm0.04$                    \\ 
                     & 4000    & 44        &  $97\pm2 $             &  $92\pm2 $                    & $2.23\pm0.06$                    \\ 
                     & 6000    & 66        &  $101\pm2$             &  $126\pm3$                    & $2.11\pm0.06$                    \\ 
                     & 8000    & 89        &  $103\pm2$             &  $159\pm2$                    & $2.05\pm0.04$                    \\ 
\hline
\multirow{3}{*}{0.8} & 4000    & 83        &  $90\pm1 $             &  $100\pm2$                    & $2.26\pm0.04$                    \\ 
                     & 6000    & 108       &  $93\pm2 $             &  $138\pm3$                    & $2.14\pm0.06$                    \\ 
                     & 8000    & 134       &  $97\pm1 $             &  $170\pm2$                    & $2.07\pm0.04$                    \\ 
\hline
\multirow{5}{*}{0.7} & 2000    & 118       &  $80\pm2 $             &  $46\pm2 $                    & $1.82\pm0.08$                    \\ 
                     & 3000    & 136       &  $81\pm1 $             &  $77\pm2 $                    & $2.08\pm0.05$                    \\ 
                     & 4000    & 152       &  $82\pm1 $             &  $104\pm1$                    & $2.13\pm0.04$                    \\ 
                     & 6000    & 184       &  $84\pm1 $             &  $150\pm3$                    & $2.09\pm0.04$                    \\ 
                     & 8000    & 214       &  $85\pm2 $             &  $195\pm2$                    & $2.08\pm0.05$                    \\ 
\hline
\multirow{4}{*}{0.65}& 3000    & 189       &  $77\pm2 $             &  $82\pm2 $                    & $2.10\pm0.07$                    \\ 
                     & 4000    & 209       &  $78\pm2 $             &  $111\pm2$                    & $2.16\pm0.06$                    \\ 
                     & 6000    & 244       &  $80\pm1 $             &  $161\pm3$                    & $2.13\pm0.05$                    \\ 
                     & 8000    & 277       &  $82\pm2 $             &  $204\pm4$                    & $2.10\pm0.07$                    \\ 
\hline
\multirow{3}{*}{0.6} & 4000    & 288       &  $75\pm2 $             &  $119\pm2$                    & $2.21\pm0.06$                    \\ 
                     & 6000    & 328       &  $75\pm1 $             &  $174\pm4$                    & $2.18\pm0.06$                    \\ 
                     & 8000    & 365       &  $77\pm2 $             &  $220\pm6$                    & $2.12\pm0.07$                    \\ 
\end{tabular}
\end{ruledtabular}
\end{table}

\begin{table}[htbp]
    \caption{Calculated values for pressure, electrical resistivity, thermal conductivity and Lorenz number of Fe$_{3}$S from first principles
    computations, with reference volume $V_0=11.82$ \AA$^3$/atom. Uncertainties of $P$ due to the equation of state fit are below 1 GPa.}
    \label{tab:cond3}
\centering
\begin{ruledtabular}
\begin{tabular}{cccccc}
    $V/V_0$                  & $T$ [K] & $P$ [GPa] & $\rho$ [$\mu\Omega$cm] & $\lambda_{th}$ [Wm$^{-1}$K$^{-1}$] & $L$ [10$^{-8}$W$\Omega$K$^{-2}$] \\
\hline
    \multirow{3}{*}{1.0} & 4000    & 25        &  $113\pm3$             &  $70\pm2 $                    & $1.96\pm0.07$                    \\
                         & 6000    & 41        &  $118\pm2$             &  $97\pm2 $                    & $1.91\pm0.06$                    \\
                         & 8000    & 59        &  $124\pm3$             &  $121\pm3$                    & $1.87\pm0.06$                    \\
\hline
    \multirow{4}{*}{0.9} & 2000    & 27        &  $111\pm3$             &  $40\pm1 $                    & $2.20\pm0.08$                    \\ 
                         & 4000    & 46        &  $109\pm4$             &  $81\pm2 $                    & $2.22\pm0.10$                    \\ 
                         & 6000    & 66        &  $113\pm2$             &  $111\pm1$                    & $2.09\pm0.04$                    \\ 
                         & 8000    & 85        &  $117\pm1$             &  $133\pm2$                    & $1.95\pm0.03$                    \\ 
\hline
    \multirow{4}{*}{0.8} & 3000    & 73        &  $108\pm3$             &  $64\pm2 $                    & $2.29\pm0.09$                    \\ 
                         & 4000    & 85        &  $107\pm4$             &  $85\pm2 $                    & $2.29\pm0.10$                    \\ 
                         & 6000    & 107       &  $107\pm2$             &  $121\pm2$                    & $2.14\pm0.06$                    \\ 
                         & 8000    & 129       &  $108\pm2$             &  $151\pm3$                    & $2.03\pm0.05$                    \\ 
\hline
    \multirow{5}{*}{0.7} & 2000    & 126       &  $103\pm4$             &  $34\pm1 $                    & $1.75\pm0.10$                    \\ 
                         & 3000    & 141       &  $102\pm2$             &  $59\pm1 $                    & $2.01\pm0.06$                    \\ 
                         & 4000    & 155       &  $101\pm2$             &  $85\pm3 $                    & $2.15\pm0.08$                    \\ 
                         & 6000    & 181       &  $100\pm3$             &  $129\pm2$                    & $2.15\pm0.07$                    \\ 
                         & 8000    & 206       &  $100\pm2$             &  $170\pm4$                    & $2.13\pm0.07$                    \\ 
\hline
    \multirow{4}{*}{0.65}& 3000    & 196       &  $102\pm2$             &  $62\pm2 $                    & $2.08\pm0.08$                    \\ 
                         & 4000    & 211       &  $99\pm2 $             &  $86\pm2 $                    & $2.15\pm0.07$                    \\ 
                         & 6000    & 240       &  $96\pm2 $             &  $136\pm3$                    & $2.17\pm0.07$                    \\ 
                         & 8000    & 267       &  $96\pm3 $             &  $178\pm5$                    & $2.14\pm0.08$                    \\ 
\hline
    \multirow{3}{*}{0.6} & 4000    & 291       &  $96\pm2 $             &  $91\pm4 $                    & $2.18\pm0.10$                    \\ 
                         & 6000    & 323       &  $95\pm2 $             &  $141\pm3$                    & $2.23\pm0.07$                    \\ 
                         & 8000    & 352       &  $91\pm2 $             &  $189\pm3$                    & $2.16\pm0.06$                    \\ 
\end{tabular}
\end{ruledtabular}
\end{table}

\begin{table}[htbp]
    \caption{Parameters of the modified thermodynamic model by \textit{de Koker and Stixrude}$^{62}$\footnote[0]{$^{62}$N. de Koker and L. Stixrude, Geophys. J. Int. \textbf{178}, 162 (2009)} for $V_0=11.82$ \AA$^3$/atom and $T_0=2000$ K. Values for extensive variables are per mol of formula units.}
\label{tab:EOS}
\centering
\begin{ruledtabular}
\begin{tabular}{llccc}
                                                                          &                 & Fe                   & Fe$_7$S              & Fe$_3$S              \\
    \hline
    $P_{xs0}$                                                             & [GPa]           & -2.335               &  0.846               &  5.534               \\
    $K_{T,xs0}$                                                           & [GPa]           &  131.4               &  137.8               &  140.0               \\
    $K_{T,xs0}^\prime$                                                    &                 &  5.161               &  4.694               &  4.736               \\
    $\alpha K_{T,xs0}$                                                    & [GPa/K]         &  8.822$\cdot10^{-3}$ &  8.620$\cdot10^{-3}$ &  7.194$\cdot10^{-3}$ \\
    $V_0\left(\frac{\partial\alpha K_T}{\partial V}\right)_{T,xs0}$       & [GPa/K]         & -1.563$\cdot10^{-2}$ & -1.660$\cdot10^{-2}$ & -1.327$\cdot10^{-2}$ \\
    $T_0\left(\frac{\partial\alpha K_T}{\partial T}\right)_{T,xs0}$       & [GPa/K]         & -3.348$\cdot10^{-3}$ & -2.376$\cdot10^{-3}$ & -1.808$\cdot10^{-3}$ \\
    $V_0^2\left(\frac{\partial^2\alpha K_T}{\partial V^2}\right)_{T,xs0}$ & [GPa/K]         &  2.840$\cdot10^{-2}$ &  5.115$\cdot10^{-2}$ &  3.534$\cdot10^{-2}$ \\
    $C_{V,xs0}$                                                           & [J/(mol K)]     &  18.50               &  185.1               &  92.90               \\
    $V_0\left(\frac{\partial C_V}{\partial V}\right)_{T,xs0}$             & [kJ/(mol K)]    &  15.84               &  317.1               &  79.57               \\
    $V_0^2\left(\frac{\partial^2 C_V}{\partial V^2}\right)_{T,xs0}$       & [kJ/(mol K)]    &  2.113$\cdot10^{-2}$ &  3.094$\cdot10^{-1}$ &  1.133$\cdot10^{-1}$ \\
    $\zeta_0$                                                             & [J/(mol K$^2$)] &  3.486               &  30.04               &  12.63               \\
    $\xi$                                                                 &                 &  0.843               &  1.096               &  1.006               \\
\end{tabular}
\end{ruledtabular}
\end{table}

\bibliography{bibfile}